\documentclass[journal]{IEEEtran}

\usepackage{cite}

\usepackage{graphicx}
\usepackage{psfrag}
\DeclareGraphicsExtensions{.xps}
\usepackage{epstopdf} 
\usepackage{subcaption}
\usepackage[cmex10]{amsmath}
\usepackage{amssymb}
\usepackage{color}
\usepackage{amsthm}

\usepackage{multirow}
\usepackage[dvipsnames,svgnames]{xcolor}
\usepackage[textwidth=30mm]{todonotes}

\DeclareMathOperator*{\argmin}{arg\,min}
\DeclareMathOperator*{\argmax}{arg\,max}

\usepackage{algorithm}
\usepackage{algpseudocode}
\algrenewcommand{\algorithmiccomment}[1]{\hskip3em$\bullet$ #1} 
\newlength\myindent
\begin{document}

\title{Adversarial Attacks on Deep-Learning Based Radio Signal Classification}

\author{Meysam Sadeghi and Erik G. Larsson
	\thanks{ \newline \indent The authors (\{meysam.sadeghi,erik.g.larsson\}@liu.se) are with Department of Electrical Engineering (ISY), Link\"{o}ping University, Link\"{o}ping, Sweden.} \vspace{-1.2em}}

\maketitle

\begin{abstract}
Deep learning (DL), despite its enormous success in many computer vision and language processing applications, is
exceedingly vulnerable to adversarial attacks.
We consider the use of DL for radio signal (modulation) classification tasks,
and present practical methods for the crafting of white-box and universal black-box adversarial attacks in that application.
We show that these attacks can considerably reduce the classification performance, with extremely small perturbations of the input. 
In particular, these attacks are significantly more powerful than classical jamming attacks,
which raises significant security and robustness concerns in the use of DL-based algorithms for the wireless physical layer.
\end{abstract}

\begin{IEEEkeywords}
	Adversarial attacks, Deep learning, Wireless security, Modulation classification, Neural networks.
\end{IEEEkeywords}

\section{Introduction}
Deep learning (DL), implemented through deep neural networks (DNNs), represents a machine-learning paradigm that has been extremely successful in the last decade,
especially in  computer vision and natural language processing applications \cite{lecun2015deep}.  This revolution has also sparked interest in applying DL in many other disciplines, including  algorithm design for wireless communication systems \cite{zhang2018deep,liang_BPCNN,sun2017learning,convnetmodrec,JSTSP_ModCls}. For example, \cite{liang_BPCNN} uses a convolutional neural network (CNN) for channel decoding, \cite{sun2017learning} studies DL-based wireless resource allocation, and \cite{JSTSP_ModCls, convnetmodrec} use DL for the classical task of radio signal (modulation) classification. Promising performance have been achieved by   DL-methods in these applications.

It has been shown that DNNs are highly vulnerable to adversarial examples, which raises major security and robustness concerns \cite{szegedy2013intriguing}. Adversarial examples are malicious inputs  that are obtained by slightly perturbing an original input, in such a way that the DL algorithm misclassifies them \cite{szegedy2013intriguing,goodfellow2014explaining}. These perturbations are not ``random white noise'', but   rather well-sought directions in the feature space that cause erroneous model outputs.

In this paper, we consider the use of DL algorithms applied to the radio signal (modulation) classification problem of \cite{convnetmodrec}, and show that this class of algorithms is extremely vulnerable to adversarial attacks. For the sake of reproducibility and cultivation of future research on this topic, we use  the publicly available GNU radio machine learning dataset of \cite{rml_datasets}. Our specific contributions are as follows. First, we present a new algorithm for generation of fine-grained white-box input-specific adversarial attacks. Second, we propose a computationally efficient algorithm for crafting white-box universal adversarial perturbations (UAP). Third, we show how one can create black-box UAP attacks. Fourth, we reveal the shift invariant property of UAPs.

\section{Brief Review of Adversarial Attacks}
We denote a DNN classifier by $f(.; \boldsymbol{\theta}): \mathcal{X} \to \mathbb{R}^C$, where $\boldsymbol{\theta}$ is the set of model parameters, $\mathcal{X}~\subset~\mathbb{R}^p$ is the input domain with $p$ being the dimension of the inputs, and $C$ is the number of classes.\footnote{\textit{Notations}: Scalars are denoted by lower case letters whereas boldface lower (upper) case letters are used for vectors (matrices). We denote by $\mathbf{I}_{N}$ the identity matrix of size $N$ and represent the $n$ column of $\mathbf{I}_{N}$ as $\mathbf{e}_{n}$.}  For every input $\mathbf{x} \in \mathcal{X}$ the classifier assigns a label $ \hat{l}(\mathbf{x} ,\boldsymbol{\theta}) = \argmax_{k} f_{k}(\mathbf{x},\boldsymbol{\theta})$ where $f_{k}(\mathbf{x},\boldsymbol{\theta})$ is the output of $f$ corresponding to the $k$th class. Given these definitions, the adversarial perturbation for input $\mathbf{x}$ and classifier $f$ is denoted by $\mathbf{r}_{\mathbf{x}}$ and is obtained as follows \cite{szegedy2013intriguing}
\begin{align}
\label{adv_eq}
	\argmin_{\mathbf{r}_{\mathbf{x}}} \quad  & \Vert \mathbf{r}_{\mathbf{x}} \Vert_{2} \\
 	s.t. \quad & \hat{l}(\mathbf{x},\boldsymbol{\theta}) \neq \hat{l}(\mathbf{x} + \mathbf{r}_{\mathbf{x}},\boldsymbol{\theta}) \notag \text{\; and \;} \mathbf{x} + \mathbf{r}_{\mathbf{x}} \in \mathcal{X}. \notag
\end{align} 
Note that $\mathbf{r}_{\mathbf{x}}$ might not be unique and we might use other norms, e.g., infinity norm. In the context of wireless communication, the $l_2$-norm is a natural choice as it accounts for the perturbation power.

In practice solving \eqref{adv_eq} is difficult, hence different suboptimal methods have been proposed to approximate the adversarial perturbation \cite{szegedy2013intriguing, goodfellow2014explaining}. Among these methods, the class of fast gradient methods (FGM) is a commonly used approach \cite{goodfellow2014explaining}. They provide computationally efficient methods for crafting adversarial examples, at the cost of coarse-grained perturbations \cite{szegedy2013intriguing}. Denoting the loss function of the model by $L(\boldsymbol{\theta}, \mathbf{x}, \mathbf{y})$, where $\mathbf{y} \in \{0, 1\}^C$ is the label vector, FGM linearizes the loss function in a neighborhood of $\mathbf{x}$, and then optimizes this linearized function. There are two variants of FGM, targeted FGM and non-targeted FGM. 

In a targeted FGM attack, the adversary is searching for a perturbation that causes the classifier to have a specific misclassification, e.g., the classifier classifies QPSK modulation as AM-DSB modulation. Therefore, denoting the one-hot encoded desired target class as $\mathbf{y}^{\text{target}}$, in targeted FGM we want to minimize $L(\boldsymbol{\theta}, \mathbf{x} + \mathbf{r}_{\mathbf{x}}, \mathbf{y}^{\text{target}})$ with respect to $\mathbf{r}_{\mathbf{x}}$. Hence, FGM linearizes the loss function as $L(\boldsymbol{\theta}, \mathbf{x} \! +\! \mathbf{r}_{\mathbf{x}}, \mathbf{y}^{\text{target}}) \approx  	L(\boldsymbol{\theta}, \mathbf{x}, \mathbf{y}^{\text{target}}) \! +   \mathbf{r}_{\mathbf{x}}^{T}  \nabla_{\!\! \mathbf{x}} L(\boldsymbol{\theta}, \mathbf{x}, \mathbf{y}^{\text{target}})$ and then minimizes it by setting $ \mathbf{r}_{\mathbf{x}} = - \alpha \; \nabla_{\!\! \mathbf{x}} L(\boldsymbol{\theta}, \mathbf{x}, \mathbf{y}^{\text{target}})$, where $\alpha$ is a scaling factor to adjust the adversarial perturbation power.

In a non-targeted FGM attack, the adversary is searching for a perturbation that causes any misclassification, i.e. the adversary is not interested in a specific misclassification and any misclassification is allowed. In a non-targeted FGM attack the loss is  $L(\boldsymbol{\theta}, \mathbf{x} + \mathbf{r}_{\mathbf{x}}, \mathbf{y}^{\text{true}})$ where $\mathbf{y}^{\text{true}}$ is the true label of $\mathbf{x}$.  FGM linearizes the loss as $L(\boldsymbol{\theta}, \mathbf{x} + \mathbf{r}_{\mathbf{x}}, \mathbf{y}^{\text{true}}) \approx  	L(\boldsymbol{\theta}, \mathbf{x}, \mathbf{y}^{\text{true}}) \! +   \mathbf{r}_{\mathbf{x}}^{T}  \nabla_{\!\! \mathbf{x}} L(\boldsymbol{\theta}, \mathbf{x}, \mathbf{y}^{\text{true}})$ and then maximizes it by setting $\mathbf{r}_{\mathbf{x}} =  \alpha \; \nabla_{\!\! \mathbf{x}} L(\boldsymbol{\theta}, \mathbf{x}, \mathbf{y}^{\text{true}})$.

Besides the targeted and non-targeted categories, the adversarial attacks can be categorized along other dimensions \cite{szegedy2013intriguing, goodfellow2014explaining}. The adversarial attacks can be divided into white-box and black-box attacks, based on the amount of knowledge that the adversary has about the model. In white-box attacks, the adversary has the full knowledge of the classifier, while in black-box attacks the adversary does not have any knowledge (or has limited knowledge) of the classifier. Adversarial attacks can also be classified based on their scope to the individual or universal attacks, which will be detailed in Section V.

\vspace{-.8em}

\section{The GNU Radio ML Dataset and Its DNN}
To study the robustness and security issues of DL-based wireless systems, we will use the GNU radio ML dataset RML2016.10a \cite{rml_datasets} and its associated DNN \cite{convnetmodrec}. The main reason behind this choice is that the dataset and the source code for its associated DNN classifier \cite{rml_datasets} are publicly available at \cite{deepsig}. 

\begin{figure}[]
	\centering
	\includegraphics[width=0.92\columnwidth, trim={7cm 7.5cm 6cm 7.8cm},clip]{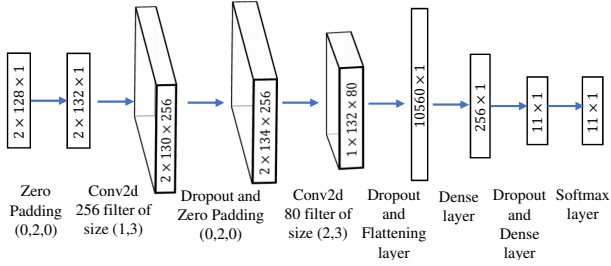}
	\caption{An illustration of VT-CNN2 of \cite{rml_datasets}.}
	\label{fig1}
	\vspace{-1.4em}
\end{figure}

The GNU radio ML dataset RML2016.10a contains $220000$ input samples, where each sample is associated with one specific modulation scheme at a specific signal-to-noise ratio (SNR). It contains $11$ different modulations, which are BPSK, QPSK, 8PSK, QAM16, QAM64, CPFSK, GFSK, PAM4, WBFM, AM-SSB, and AM-DSB. The samples are generated for $20$ different SNR levels from $-20$ dB to $18$ dB with a step of $2$ dB. Each sample input is a vector of size $256$, which corresponds to $128$ in-phase and $128$ quadrature components. Half of the samples are considered as the training set and the other half as the test set. \cite{rml_datasets} uses a deep CNN classifier named as VT-CNN2. The structure of VT-CNN2 is illustrated in Fig.~\ref{fig1}, following TensorFlow's default format for data, i.e., (height, width, channels). We use this network in our analysis.  

\vspace{-0.5em}

\vspace{-0.4em}

\section{Adversarial Attacks for DL-based Modulation Classification}

In this section, we develop a white-box  adversarial attack on DL-based  modulation classification, using VT-CNN2 as the classifier. 
(A black-box attack is devised in Section~\ref{sec5}.) In a wireless system, when the attacker is absent, the receiver (RX) receives a wireless signal from one (or multiple) legitimate transmitter (TX), which is denoted by $\mathbf{x}$. But when the attacker is present, it also transmits a signal to create a low power perturbation $\mathbf{r}_{\mathbf{x}}$ at the RX. Therefore, the RX will receive $\mathbf{x}_{\text{adv}} = \mathbf{x} + \mathbf{r}_{\mathbf{x}}$. The attacker target is to design $\mathbf{r}_{\mathbf{x}}$ such that it causes misclassification for the underlying DNN at the RX side.

To design an adversarial perturbation $\mathbf{r}_{\mathbf{x}}$ for a given input $\mathbf{x}$, we start with the white-box attack for simplicity. Later in Section V, we extend the attack to more general cases. FGMs are computationally efficient methods for crafting adversarial perturbations, but they provide coarse-grained perturbations and also have a low success rate for fooling the classifier. Therefore, we present Alg.~\ref{alg1} to address these issues. 

\begin{algorithm}[]
	\caption{Crafting an adversarial example \label{alg1}}
	\begin{algorithmic}[1]
		\Statex Inputs: \vspace{-0.5em}
		\begin{itemize}
			\item input $\mathbf{x}$ and its label $l_{true}$ \vspace{-0.2em}
			\item the model $f(.,\boldsymbol{\theta})$ \vspace{-0.2em}
			\item desired perturbation accuracy $\varepsilon_{acc}$  \vspace{-0.2em}
			\item maximum allowed perturbation norm $p_{max}$  \vspace{-0.2em}
		\end{itemize} 
		\Statex Output: adversarial perturbation of the input, i.e., $\mathbf{r}_{\mathbf{x}}$ \vspace{-0.3em}
	\vspace{-0.5em} \noindent\rule{8.2cm}{0.2pt}
		\State Initialize: $\boldsymbol{\varepsilon} \leftarrow \mathbf{0}^{C \times 1} $ 
		 
		\For{\textit{class-index} in \textit{range}($C$)} \vspace{-0.2em} 
		
			\State $\varepsilon_{max} \leftarrow p_{max}$, $\varepsilon_{min} \leftarrow 0$
			
			\State $\mathbf{r}_{norm} =  \left(\Vert \nabla_{\!\! \mathbf{x}} L(\mathbf{x}, \mathbf{e}_{\text{\textit{class-index}}}) \Vert_{2} \right)^{-1} \nabla_{\!\! \mathbf{x}} L(\mathbf{x}, \mathbf{e}_{\text{\textit{class-index}}})$ 
			
			\While{$\varepsilon_{max} - \varepsilon_{min} > \varepsilon_{acc}$}
			
				\State $\varepsilon_{ave} \leftarrow (\varepsilon_{max} + \varepsilon_{min})/2 $
				
				\State $\mathbf{x}_{adv} \leftarrow  \mathbf{x} - \varepsilon_{ave} \; \mathbf{r}_{norm}$
				
				\If{$\hat{l}(\mathbf{x}_{adv}) == l_{true}$} \vspace{-0.2em}
					\State $\varepsilon_{min} \leftarrow \varepsilon_{ave}$ \vspace{-0.3em}
				\Else  \vspace{-0.3em}
					\State $\varepsilon_{max} \leftarrow \varepsilon_{ave}$
				\EndIf
				
			\EndWhile
			
			\State $\left[ \boldsymbol{\varepsilon} \right]_{\text{\textit{class-index}} }=  \varepsilon_{max}$
			
		\EndFor
		
		\State $ \text{\textit{target-class}} = \argmin \; {\boldsymbol{\varepsilon}}$     and     $\varepsilon^* = \min \; {\boldsymbol{\varepsilon}}$

		\State $\mathbf{r}_{\mathbf{x}} = - \dfrac{\varepsilon^*}{\Vert \nabla_{\!\! \mathbf{x}} L(\mathbf{x}, \mathbf{e}_{\text{\textit{target-class}}}) \Vert_{2}}    \nabla_{\!\! \mathbf{x}} L(\mathbf{x}, \mathbf{e}_{\text{\textit{target-class}}})$ 

	\end{algorithmic}
\end{algorithm}

Alg.~\ref{alg1} improves two specific drawbacks of FGM. First, FGM is designed to set the scaling factor of the perturbation, i.e., $\alpha$, such that it goes all the way to the edge of a norm ball surrounding the input \cite{goodfellow2014explaining}. However, Alg.~\ref{alg1} uses a bisection search to find the exact value of scaling factor that guarantees the misclassification (within the extent of the constraint on the perturbation norm). Second, in a non-targeted FGM attack, FGM tries to increase $L(\boldsymbol{\theta}, \mathbf{x} \! +\! \mathbf{r}_{\mathbf{x}}, \mathbf{y}^{\text{true}})$, and for a targeted attack FGM tries to minimize $L(\boldsymbol{\theta}, \mathbf{x} \! +\! \mathbf{r}_{\mathbf{x}}, \mathbf{y}^{\text{target}})$ just for a specific target class. On the contrary, Alg.~\ref{alg1} searches among all possible targeted attacks and then select the one with the least perturbation required to enforce misclassification. Therefore, Alg.~\ref{alg1} provides fine-grained adversarial perturbations while relying on the computationally efficient FGM as the core of the algorithm.

In the computer vision literature on adversarial attacks, the focus is on finding slight perturbations that a human observer does not even notice,
 while it causes misclassification. Given Alg.~\ref{alg1}, one can think of a similar analogy in wireless applications, perturbations which are unnoticeable (or quasi-unnoticeable) by the receiver. Here we propose  two new metrics, the perturbation-to-noise ratio (PNR) and the perturbation-to-signal ratio (PSR), where PNR is the ratio of the perturbation power to the noise power and PSR is the ratio of the perturbation power to the signal power. Note that the signal-to-noise ratio (SNR) is related to PSR and PNR as $\text{PNR} = \text{PSR} \times \text{SNR}$ or equivalently $\text{PNR [dB]} = \text{PSR [dB]} + \text{SNR [dB]}$. Given these definitions, we can consider a perturbation (quasi) imperceptible if for that perturbation we have $ \text{PNR} \leq 1$, as the perturbation will be in the same order or even below the noise level. 

\begin{figure}[]
	\centering
	\includegraphics[width=0.95\columnwidth, trim={2.3cm 7.5cm 2.5cm 8cm},clip]{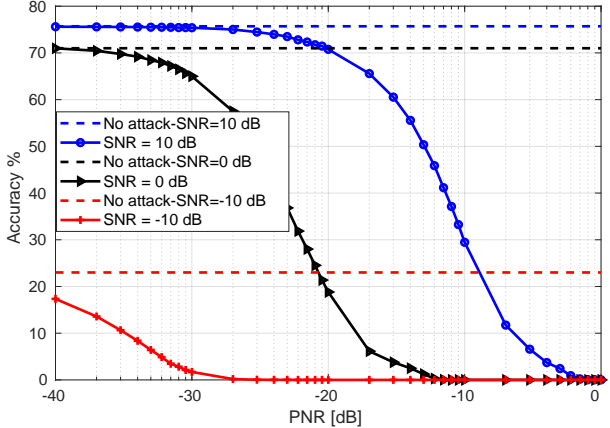}  
	\caption{The accuracy of VT-CNN2 versus PNR, with and without adversarial attack.}
	\label{fig2}
	\vspace{-1.5em}
\end{figure}

Fig. \ref{fig2} presents the accuracy of VT-CNN2 versus PNR, for three different values of SNR. The perturbations are created using Alg.~\ref{alg1}. The horizontal dashed lines represent the accuracy of VT-CNN2 when there is no attack. From Fig.~\ref{fig2}, it is obvious that when the perturbation is in the same order as the noise (for all three SNR levels), the attack can cause $100\%$ misclassification. Note that, even when the perturbation is one or several orders of magnitude less than the noise level, the attack can significantly reduce the accuracy of the model. This raises a major concern regarding the robustness of DL-based wireless application and reveals their vulnerability to white-box adversarial attacks.

\section{Universal Black-box Attacks for Wireless Communication Systems}
\label{sec5}
\vspace{-0.1em}
In the previous section, we presented a white-box attack while considering three limiting assumptions. First, the attacker knows the exact input. Second, each element of $\mathbf{x}$ is perturbed by its corresponding element in $\mathbf{r}_{\mathbf{x}}$, i.e., the attacker is synchronous with the transmitter. Third, as we considered a white-box attack, we assumed the attacker has a perfect knowledge of the underlying model, i.e., $f(. ; \boldsymbol{\theta})$. In this section, we address these limiting assumptions.

\vspace{-0.1em}

\subsection{Universal Adversarial Perturbations}
Alg.~\ref{alg1} creates input-dependent adversarial perturbations, i.e., given input $\mathbf{x}$ it generates a perturbation $\mathbf{r}_{\mathbf{x}}$ to fool the model. This enforces the attacker to know the input of the model, which is not a practical assumption. Therefore, it is interesting to create adversarial attacks which are input-agnostic. More precisely, instead of $\mathbf{r}_{\mathbf{x}}$, we are interested to find a universal adversarial perturbation $\mathbf{r}$ that can fool the model with high probability, independent of the input applied to the model. In the literature on ML and computer vision, such a perturbation is called a UAP \cite{moosavi2017universal}. 

A common method for creating UAP is presented in \cite{moosavi2017universal}. The algorithm therein, receives as inputs, 1) the model, 2) the desired norm of the UAP, and 3) a random subset of data inputs, e.g., $\{ \mathbf{x}_{1}, \ldots, \mathbf{x}_{N} \}$. 
Based on these inputs, it generates as output a UAP $\mathbf{r}$. The core of the algorithm is an iterative approach that in each iteration requires to generate an adversarial perturbation for each of the $N$ data points, e.g., by running Alg.~\ref{alg1} $N$ times. Hence, it is computationally expensive.

In this section, we propose a new algorithm for generating a UAP that has a very low computational complexity and also provides a better fooling rate on our dataset compared to \cite{moosavi2017universal}. The algorithm uses principal component analysis (PCA) to craft the UAP. The main intuition behind the algorithm is as follows. Assume we have an arbitrary subset of inputs $\{ \mathbf{x}_{1}, \ldots, \mathbf{x}_{N} \}$, and their associated perturbation directions $\{ \mathbf{n}_{\mathbf{x}_{1}}, \ldots, \mathbf{n}_{\mathbf{x}_{N}} \}$, where $\mathbf{n}_{\mathbf{x}_{i}} = \nabla_{\!\! \mathbf{x}_{i}} L(\boldsymbol{\theta}, \mathbf{x}_{i}, \mathbf{y}^{\text{true}}) / \Vert \nabla_{\!\! \mathbf{x}_{i}} L(\boldsymbol{\theta}, \mathbf{x}_{i}, \mathbf{y}^{\text{true}}) \Vert_{2}$. Now the question is, how one can craft a UAP $\mathbf{r}$ that contains the common characteristic(s) of $\{ \mathbf{n}_{\mathbf{x}_{1}}, \ldots, \mathbf{n}_{\mathbf{x}_{N}} \}$? Noting that $\mathbf{n}_{\mathbf{x}_{1}}$ to $\mathbf{n}_{\mathbf{x}_{N}}$ are points in $\mathbb{R}^{p}$, if we stack them into a matrix, then the first principal component of the matrix would have the largest variance. In other words, the first principal component will account for as much as variability in $\{ \mathbf{n}_{\mathbf{x}_{1}}, \ldots, \mathbf{n}_{\mathbf{x}_{N}} \}$ as possible. Therefore, we suggest using the direction of the first principal component as the direction of UAP. The detailed algorithm is given in Alg.~\ref{alg2}.

\begin{algorithm}[]
	\caption{PCA-based approach for crafting a UAP \label{alg2}}
	\begin{algorithmic}[1]
		\Statex Inputs: \vspace{-0.5em}\begin{itemize}
			\item a random subset of input data points $\{ \mathbf{x}_{1}, \ldots, \mathbf{x}_{N} \}$ and their corresponding labels \vspace{-0.2em}
			\item the model $f(.,\boldsymbol{\theta})$ \vspace{-0.2em}
			\item maximum allowed perturbation norm $p_{max}$  \vspace{-0.2em}
		\end{itemize} 
		\Statex Output: a UAP $\mathbf{r}$ \vspace{-0.5em}
		
		\vspace{-0.1em} \noindent\rule{8.2cm}{0.3pt}
		\State Evaluate $\mathbf{X}^{N \times p}=[ \mathbf{n}_{\mathbf{x}_{1}}, \ldots, \mathbf{n}_{\mathbf{x}_{N}}]^{T}$. 
		\State Compute the first principal direction of $\mathbf{X}$ and denote it by $\mathbf{v}_{1}$, i.e., $\mathbf{X} =  \mathbf{U} 	\mathbf{\Sigma} \mathbf{V}^{T}$ and $\mathbf{v}_1 = \mathbf{V} \; \mathbf{e}_1$.
		\State $\mathbf{r} = p_{max} \mathbf{v}_1$.
	\end{algorithmic}
\end{algorithm}

\begin{figure}[]
	\centering
	\includegraphics[width=1.1\columnwidth, trim={0.3cm 0.2cm 0cm 1.2cm},clip]{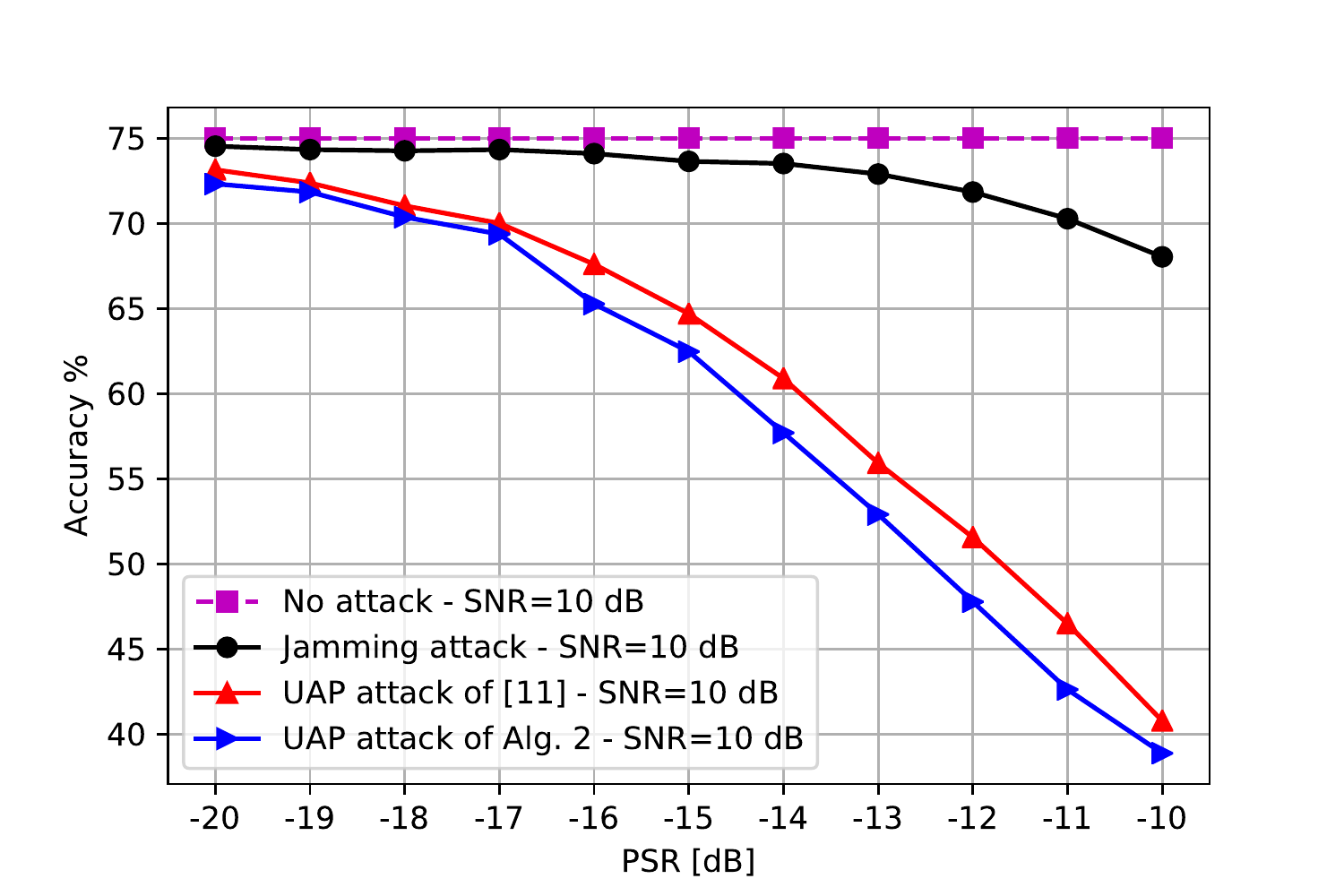}
	\caption{The accuracy of VT-CNN2 under different attacks.}
	\label{fig3}
	\vspace{-1.5em}
\end{figure}

Fig. \ref{fig3} investigates the performance of Alg.~\ref{alg2}. It illustrates the accuracy of VT-CNN2 versus PSR, for our proposed UAP attack, the UAP attack presented in \cite{moosavi2017universal}, and a jamming attack. For the jamming attack, the adversary creates Gaussian noise, which has the same mean as the data points and same power as the UAP attacks. Note that Alg.~\ref{alg2} provides higher fooling rate than \cite{moosavi2017universal}. Moreover, even for very small PSR values the performance of VT-CNN2 drops significantly, e.g., for PSR$=-10$ dB the accuracy drops by half. Also note that the proposed UAP is significantly more powerful than the classical jamming attack.

To emphasize the low computational cost of Alg.~\ref{alg2}, we also present Table \ref{table1}, which compares the run-time of Alg.~\ref{alg2} with \cite{moosavi2017universal} in seconds, for SNR$=10$ dB and $N=50$. All the simulations are performed using TensorFlow on an NVIDIA GeForce GTX 1080 Ti graphic processing unit. Note that \cite{moosavi2017universal} requires much more time to craft a UAP as PSR reduces, while Alg.~\ref{alg2} provides a steady and efficient computational performance.

\begin{table}[]
	\centering
	\begin{tabular}{|l|l|l|l|l|l|l|}
		\hline
		PSR {[}dB{]}                                                  & -10  & -12  & -14  & -16  & -18  & -20  \\ \hline
		Time required by \cite{moosavi2017universal}                  & 20.5 & 23.0 & 25.1 & 27.2 & 29.0 & 30.5  \\ \hline
		Time required by Alg.~\ref{alg2}                              & 0.3  & 0.3  & 0.3  & 0.3  & 0.3  & 0.3 \\ \hline
	\end{tabular}
\vspace{-0.5em}
\caption{Run time  of Alg.~\ref{alg2} compared to \cite{moosavi2017universal} in seconds, for SNR$=10$ dB and $N=50$.}
\label{table1}
\end{table}

\vspace{-0.8em}

\subsection{Black-Box Attacks and Shift Invariant Property of UAPs}
In the previous subsection, 1) we assumed the attacker has the perfect knowledge of the model $f(.,\boldsymbol{\theta})$, 2) and it is synchronous with the transmitter, i.e., each element of $\mathbf{x}$ to be perturbed by its corresponding element in $\mathbf{r}$. In the following, we show how an attacker can address these two limitations.

To address the first problem we use the transferability property of adversarial examples \cite{goodfellow2014explaining}. Due to this property, an adversarial example crafted for a specific DNN can also fool other DNNs with different architectures, with high probability \cite{goodfellow2014explaining}. Therefore, to craft a UAP for VT-CNN2, we first create such a UAP for a substitute DNN and then apply it on VT-CNN2. Here we consider a $256-1024-1024-1024-512-128-11$ fully connected multilayer perceptron (MLP) as our substitute DNN and craft a UAP for it. 

To address the second problem, we reveal an interesting property of the crafted UAPs, namely, the shift invariant property. More precisely, we show that the UAPs created by Alg.~\ref{alg2} are shift invariant, i.e, any circularly shifted version of them can fool the DNN and cause misclassification.

\begin{figure}[]
	\centering
	\includegraphics[width=1.1\columnwidth, trim={0.3cm 0.22cm 0cm 1.18cm},clip]{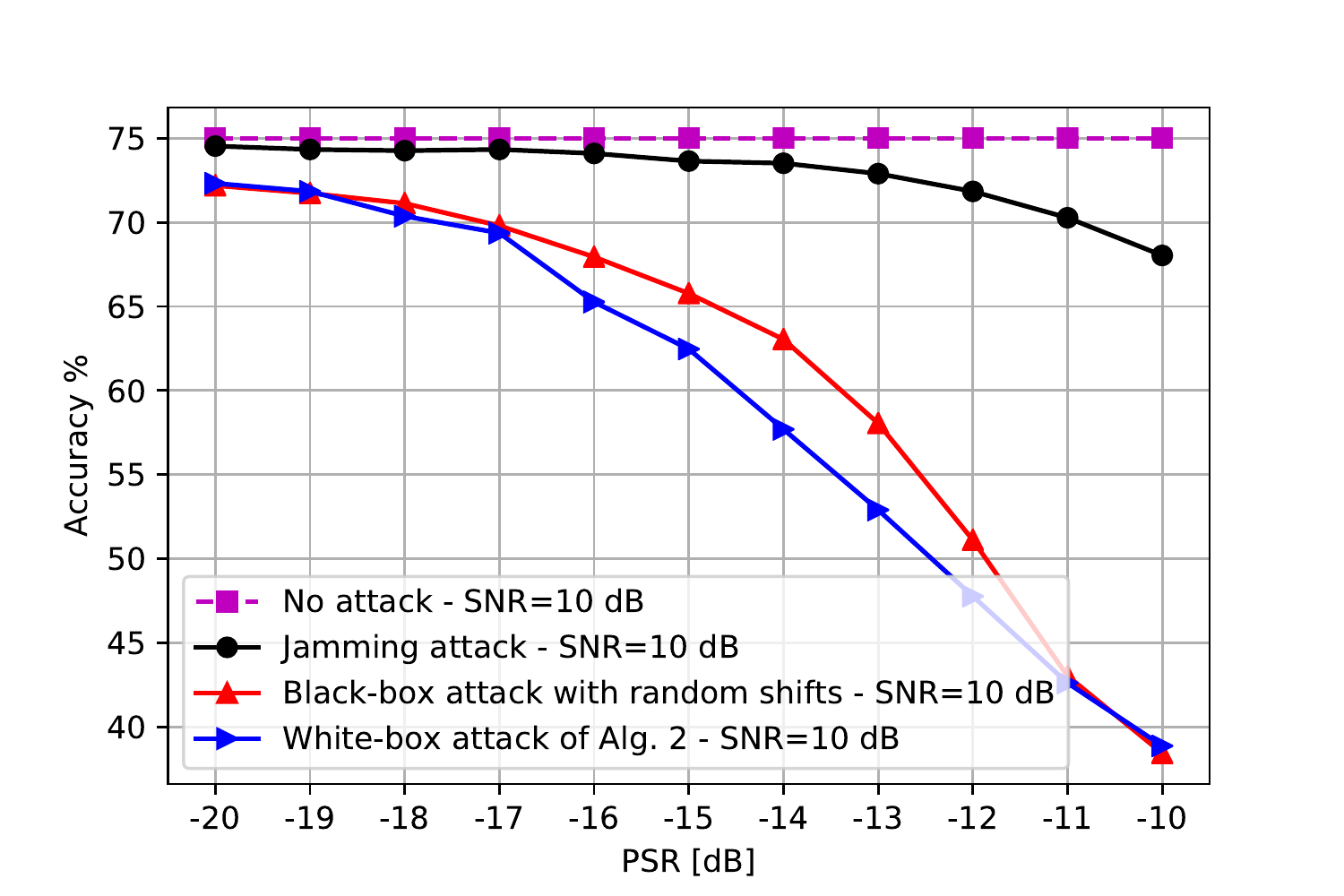}
	\caption{An illustration of transferability and shift invariant properties of the proposed UAP attack.}
	\label{fig4}
	\vspace{-1em}
\end{figure}

Fig. \ref{fig4} shows the performance of two UAP attacks designed using Alg. \ref{alg2}, a white-box UAP attack that has the perfect knowledge of the model, and a black-box UAP attack with random shifts. For the latter case the UAP is crafted for the aforementioned substitute MLP (black-box attack) and then the UAP is randomly shifted (non-synchronous attack). Given Fig. \ref{fig4}, note the following observations. First, the black-box attack is approximately as effective as the white-box attack. Second, any random shifted version of the UAP is nearly as destructive as the original synchronous version, hence there is no need for a synchronous attack. Therefore, Fig. \ref{fig4} shows that we are able to craft  extremely low power UAPs that can cause severe misclassification, while we neither need to know the model of the underlying DNN, nor require a synchronous attack.

\vspace{-1em}

\section{Conclusion}

We considered the use of DL-based algorithms for radio signal classification and showed that these algorithms are
extremely susceptible to adversarial attacks. 
Specifically we designed white-box and black-box attacks on a DL classifier and demonstrated their  effectiveness.
Significantly less transmit power is required by the attacker in order to cause misclassification, as compared to the case of
conventional jamming (where the attacker transmits only random noise). This exposes a fundamental vulnerability of
DL-based solutions.

Given the openness (broadcast nature)  of the wireless transmission medium, we conjecture that other DL-based 
signal processing algorithms for the wireless physical layer may suffer from the same security problem.

\vspace{-0.4em}

\bibliographystyle{IEEEtran}
\bibliography{IEEEabrv,DLandWC}

\end{document}